\documentclass[twocolumn,showpacs,preprintnumbers,showkeys,superscriptaddress]{revtex4}
\usepackage{graphicx}
\usepackage{dcolumn}
\usepackage{bm}

\def\eqref#1{Eq.~(\ref{eq:#1})}

\def\figlab#1{\label{fig:#1}}

\def\tablab#1{\label{tab:#1}}

\begin{document}

\title {Robustness of regularities for energy centroids  in the presence
of  random interactions}
\author{Y. M. Zhao}     \email{ymzhao@sjtu.edu.cn}
\affiliation{Department of Physics,  Shanghai Jiao Tong
University, Shanghai 200240, China} \affiliation{Center of
Theoretical Nuclear Physics, National Laboratory of Heavy Ion
Accelerator, Lanzhou 730000, China}
\affiliation{Cyclotron Center,  Institute of Physical Chemical Research (RIKEN), \\
Hirosawa 2-1, Wako-shi,  Saitama 351-0198,  Japan}
\author{A. Arima} \affiliation{Science Museum,
Japan Science Foundation, 2-1 Kitanomaru-koen, Chiyodaku, Tokyo
102-0091, Japan}
\author{N. Yoshida} \affiliation{Faculty of
Informatics, Kansai University, Takatsuki, 569-1095, Japan}
\author{K. Ogawa}
\affiliation{Department of Physics, Chiba University, Yayoi-cho
1-33, Inage, Chiba 263-8522, Japan}
\author{N. Yoshinaga}
\affiliation{Department of Physics, Saitama University, Saitama
338-8570, Japan}
\author{V.K.B. Kota} \affiliation{Physical Research Laboratory, Ahmedabad 380 009, India}

\date{\today}

\begin{abstract}
In this paper we study energy centroids such as those with fixed
spin and isospin, those with fixed irreducible representations for
both bosons and fermions, in the presence of random two-body
and/or three-body interactions. Our results show that regularities
of energy centroids of fixed spin states reported in earlier works
are very robust in these more complicated cases. We suggest that
these behaviors might be intrinsic features of quantum many-body
systems interacting by random forces.
\end{abstract}


\maketitle

\section{INTRODUCTION}

In 1998 Johnson, Bertsch and Dean obtained a preponderance of
${\rm spin}^{\rm parity}=0^+$ ground states for even-even nuclei
in the presence of random two-body interactions  \cite{Johnson}.
Since then, there have been a lot of efforts towards understanding
this observation. Studies along this line were reviewed in Refs.
\cite{Zhao-review,Zelevinsky}.

Although recently there were  progresses
\cite{Papenbrock,Yoshinaga,Otsuka} in evaluation of ground state
energies, finding the ground state by a simple approach is usually
difficult in the presence of random interactions. Thus one may
study other quantities which are relatively simple. Along this
line, much attention [7-13] has been paid to spin $I$ energy
centroids (defined by the average energy of spin $I$ states and
denoted by $\overline{E_I}$). Main results are reviewed briefly as
below.

 (1) From numerical experiments by using the TBRE, it
was found  in Ref. \cite{Zhaox}  that $\overline{E_I}$'s with
$I\simeq I_{\rm min}$ or $I\simeq I_{\rm max}$ have large
probabilities to be the lowest while those with other $I$ have
very small probabilities to be the lowest. Roughly speaking, there
are $\sim 50\%$ of the cases for which  $\overline{E_I}$ with
$I\simeq I_{\rm min}$ ($I\simeq I_{\rm max}$) is the lowest. We
define $\langle \overline{E_I} \rangle_{\rm min}$
 ($\langle \overline{E_I} \rangle_{\rm max}$) as the value
  obtained by averaging
 $\overline{E_I}$ over the subset where
 $\overline{ E_{I \simeq I_{\rm min}}}$
 ($\overline{ E_{I \simeq I_{\rm max}}}$) is the lowest energy.
Ref. \cite{Zhaox}  also demonstrated that
 $\langle \overline{E_I} \rangle_{\rm min} \simeq
CI(I+1)$ and $\langle \overline{E_I} \rangle_{\rm max} \simeq
C\left[ I_{\rm max} (I_{\rm max}+1) - I(I+1) \right]$, where the
value of coefficient $C$  depends on the active single-particle
orbits and the choice of the ensemble. We have studied cases of
single-$j$ shell configurations as well as many-$j$ shell
configurations in which shells are denoted by $j_1, j_2, \cdots$,
etc. For the TBRE,  $C\simeq 1/(4 \sum_i  j_i^2$). The
regularities of $\overline{E_I}$ were argued  in Ref. \cite{Zhaox}
by assuming that two-body coefficients of fractional parentage
(cfp's) behave like randomly, and the behavior of $\langle
\overline{E_I} \rangle_{\rm min} \simeq CI(I+1)$ was reproduced
for four fermions in a $j=17/2$ shell under this simple
assumption.

The above regularities of $\overline{E_I}$ which are stable for
single-closed shell (both single-$j$ and many-$j$ shells) were
found in Ref. \cite{Zhao-2005} to be robust even for many-$j$
shells where each orbit can have positive or negative parity and
for systems with isospins.  For cases of nucleons in many-$j$
shells (different $j$ can have different parity), the value of
coefficient $C^{\pm}$ in the relation $\langle
\overline{E_{I^{\pm}}} \rangle_{\rm min} \simeq C^{\pm}
I^{\pm}(I^{\pm}+1)$ are sensitive to the $j$ values but not to
parity or isospin. Very recently, one of the authors of the
present paper, Kota,  studied in Ref. \cite{Kota2} energy
centroids with fixed irreducible representations of some of the
group symmetries of the interacting boson models
\cite{Arima0,Iachello} such as the $sd$ interacting boson model
(IBM), the $sd$ IBM with isospin, etc. It was found that the
lowest and highest irreducible representations carry most of the
probability for the corresponding centroids to be the lowest in
energy. This is a generalization of results found numerically in
Refs. \cite{Zhaox,Zhao-2005}.

  (2)  Mulhall, Volya, and Zelevinsky assumed in Ref. \cite{Mulhall1} the geometric
chaos (quasi-randomness in the process of angular momentum
couplings)  and derived a linear relation between $\overline{E_I}$
and $I(I+1)$;   The same result was derived  in Ref. \cite{Kota}
by resorting to the group structure of $U(2j+1) \supset O(3)$ for
$n$ fermions in a single-$j$ shell. Let us define single-$j$
Hamiltonian
\begin{eqnarray}
&&  H =  -  \sum_J G_J \frac{\sqrt{2J+1}}{2} \left(   \left(
a_{j}^{\dagger}  a_{j}^{\dagger} \right)^{(J)} \times
 \left( \tilde{a}_{j}  \tilde{a}_{j} \right)^{(J)} \right)^{(0)}
 ~. \label{pair} \nonumber \\
\end{eqnarray}
The formula of $\overline{E_I}$ of Refs. \cite{Mulhall1,Kota} was
written as follows,
\begin{eqnarray}
&& \overline{E_I} = \sum_J (2J+1) G_J \left( \frac{n}{2j+1} \right)^2 \nonumber \\
& +& I (I+1) \sum_J (2J+1) \frac{3 (J(J+1) - 2 j(j+1)) }{2 j^2
(j+1)^2 (2j+1)^2}  G_J \nonumber \\
&+& O (I^2(I+1)^2), \label{chaotic}
\end{eqnarray}
where $O (I^2 (I+1)^2) $ refers to higher $I$ terms which seem
negligible. The first term of this formula is a constant
independent of $I$.  The  second term of Eq. (\ref{chaotic}) is
proportional to $I(I+1)$. Thus we have the relation
$\overline{E_I} \simeq E_0 + C I(I+1)$. However, the value of
coefficient $C$ thus obtained was found  in Ref. \cite{Zhao-2005}
to be systematically smaller than those obtained by the TBRE or
empirical formula $C\simeq 1/(4j^2)$; and furthermore, even for
systems in which one cannot assume randomness of the geometric
chaos or randomness of the cfp's, a similar pattern was found to
occur. Therefore, the arguments of Refs.
\cite{Mulhall1,Kota,Zhaox} are just part of the full story and a
sound understanding is not yet available. It is interesting to
note that one can obtain the $I(I+1)$ term of Eq. (\ref{chaotic})
with an additional factor of $\frac{1}{2}$ when one transforms the
single-$j$ hamiltonian of Eq. (\ref{pair}) into its particle-hole
form via the Pandya transformation.

The purpose of this paper is as follows. First, we  discuss energy
centroids with fixed spin $I$ and isospin $T$, denoted by
$\overline{E_{I,T}}$'s. Although  Ref. \cite{Zhao-2005} discussed
$\overline{E_I}$'s for systems with isospin,  $\overline{E_I}$'s
with different isospin $T$ were mixed.  Second, we  study a
Hamiltonian with random three-body interactions for $sd$ bosons,
while earlier works studied the property of $\overline{E_I}$'s by
using random two-body interactions.   Third, we   study
$\overline{E_{\{ f \}}}$ with a fixed irreducible representation
$\{ f \}$, as an extension of the work in Ref. \cite{Kota2}. These
results are discussed by using propagation equations.

This paper is organized as follows. In Sec. II, we discuss results
of $\overline{E_{I,T}}$'s (energy centroids of states with  given
spin $I$ and isospin $T$) and $\overline{E_T}$'s (energy centroids
of states with  given isospin $T$) for proton and neutron systems,
where one will see that $\overline{E_{I,T}}$'s are approximately
linear in terms of $I(I+1)$ and that $\overline{E_T}$ is precisely
linear in terms of $T(T+1)$. In Sec. III, we present results of
$\overline{E_I}$ with random three-body interactions for $sd$
bosons, where one sees that regularities of energy centroids of
spin $I$ states under random three-body interactions are very
similar to those under two-body interactions.  In Sec. IV, we
discuss our results of energy centroids with fixed irreducible
representations, where one will see that energy centroids with
lowest and highest irreducible representations carry most of the
probability to be the lowest. In Sec. V we discuss and summarize
the results obtained in the present work. In Appendix A we present
a few formulas which are useful in deriving propagation equations
of this paper.

In this paper we take the Two-body Random Ensemble (TBRE) for
two-body matrix elements, with the same definition given in Ref.
\cite{Zhao-review}. For three-body random interactions for $sd$
boson Hamiltonian, we take the same definition given in Ref.
\cite{Bijker}; in Appendix B we present the definition of
three-body Hamiltonian for $sd$ bosons, for the sake of
convenience.

\section{energy centroids
of  spin $I$ and isospin $T$ states}

In this Section we investigate regularities of
$\overline{E_{I,T}}$. Our $\overline{E_{I,T}}$ values are obtained
by using  $\overline{E_I}$ of  all systems with $N_p$ valence
protons and  $N_n$ valence neutrons under the requirement $N_p+
N_n = N$. We first obtain number of states with fixed $I$ and $T$,
denoted by $D_{IT}$, which is given by $D_{I}(Np=\frac{n+2T}{2},
N_n=\frac{n-2T}{2}) - D_{I}(Np=\frac{n+2T+2}{2},
N_n=\frac{n-2T-2}{2})$ where $D_I$ is the number of spin $I$
states with $T_Z=(N_p-N_n)/2 $. We denote $\overline{E_I}
(Np=\frac{n+2T}{2}, N_n=\frac{n-2T}{2})$ by
$\overline{E_I^{T_Z=T}}$, and $D_{I} (Np=\frac{n+2T}{2},
N_n=\frac{n-2T}{2})$ by $D_I^{T_Z=T}$. Using these
$\overline{E_I^{T_Z=T}}$ and  $D_I^{T_Z=T}$, we can write
$\overline{E_{I,T}}$ explicitly as follows.
\begin{eqnarray}
\overline{E_{I,T}} = \left( \overline{E_I^{T_Z=T}} \times
D_I^{T_Z=T} -\overline{E_I^{T_Z=T+1}} \times D_I^{T_Z=T+1}
\right)/D_{IT}. \nonumber
\end{eqnarray}

\begin{figure}[tbp]
\includegraphics[width=8.4cm,clip]{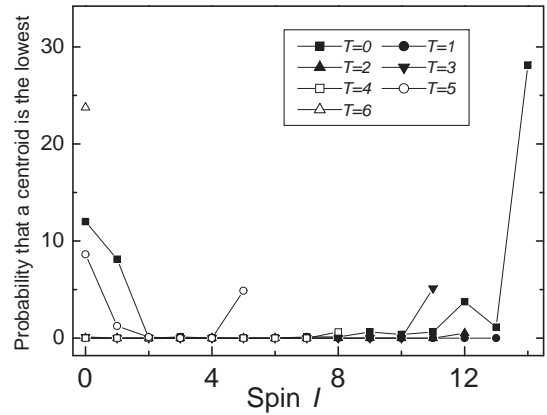}
\caption[fig1]{ Probability that an energy centroid
$\overline{E_{I,T}}$ is the lowest.  One sees that this
probability is large only when $I$ $and$ $T$ are close to minimum
or maximum. Note that the $I_{\rm max}$ value for large $T$
becomes smaller as $T$ increases, and that  for 12 nucleons in the
$sd$ shell only $I=0$ is possible when $T=6$. } \figlab{1}
\end{figure}

We have obtained $\overline{E_{I,T}}$ based on $\overline{E_I}$'s
of systems with $N=12$ ($N_p=6, 7, \cdots 12, N_p+N_n=12$) and
those with $N=8$ ($N_p=4, 5,\cdots 8, N_p+N_n=8$) in the $sd$
shell. In Fig. 1 we present out results of ${\cal P}(I,T)$ which
is the probability that the lowest energy centroid has spin $I$
and isospin $T$. One easily notices that ${\cal P}(I,T)$ is
sizable only when both $I$ and $T$ are close to their minimum or
maximum values. To see this more clearly, we list in Table I the
values of $\sum_I {\cal P}(I,T)$, where we see that $\sum_I {\cal
P}(I,T)$ is very small when the value of $T$ is not close to its
minimum or maximum value. We note here that $\sum_I {\cal P}(I,T)$
does not equal ${\cal P}(T)$, the probability that
$\overline{E_T}$ (the energy centroid of all states with isospin
$T$) is the lowest.

The feature that $\sum_I {\cal P}(I,T)$ is large only when $T$ and
$I$ are close to their minimum or maximum values is very similar
to that of ${\cal P}(I)$ discussed in Refs.
\cite{Zhaox,Zhao-2005}, and more generally,  to that of  ${\cal
P}(\lambda)$ discussed in Ref. \cite{Kota2} where $\lambda$
denotes irreducible representation of states in interacting boson
models suggested in Refs. \cite{Arima0,Iachello}.

\begin{table}
\caption{\tablab{par} $\sum_I {\cal P}(I,T)$ for eight  and twelve
particles in the $sd$ shell.  One sees that $\sum_I {\cal P}(I,T)$
is very small when  the value of $T$ is not close to $T_{\rm min}$
or $T_{\rm max}$.  We define $\overline{E_T} = \sum_{I} (2I+1)
\overline{E_{I,T}}  D_{IT}   / \left(\sum_I (2I+1) D_{IT} \right)$
and ${\cal P}(T)$ be the probability for $\overline{E_T}$ to be
the lowest in energy, and note here that ${\cal P}(T)$ does not
equal $\sum_I {\cal P}(I,T)$. }
\begin{ruledtabular}
\begin{tabular}{cccccccc}
 $T$& 0 & 1 & 2& 3 &4 &5 & 6 \\ \hline
\multicolumn{8}{c}{$N=8$}\\
& 47.1 & 11.0 & 5.5 & 2.2 & 34.2 \\ \hline
\multicolumn{8}{c}{$N=12$}\\
& 57.9 & 0.8 & 0.5 & 4.9 & 0.5 & 13.6 & 21.8 \\
\end{tabular}
\end{ruledtabular}
\end{table}

\begin{figure}[tbp]
\includegraphics[width=8.7cm,clip]{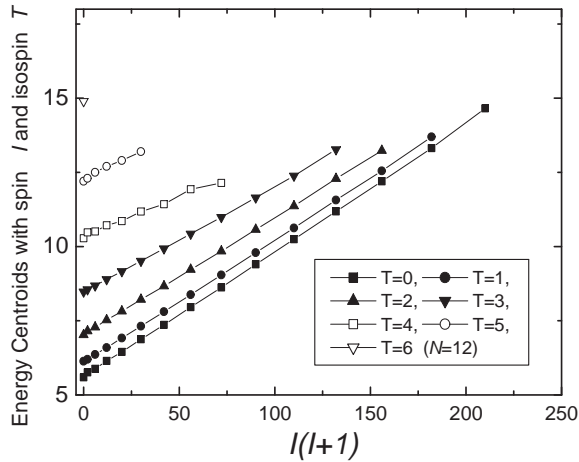}
\caption[fig2]{ $\langle \overline{E_{IT}} \rangle_{\rm min}$
versus $I(I+1)$ values.  Here $\langle \overline{E_{IT}}
\rangle_{\rm min}$ are obtained by averaging the
$\overline{E_{IT}}$ over the cases in which the lowest
$\overline{E_{IT}}$  has $I=0,1$ and $T=0$. One sees that $\langle
\overline{E_{IT}} \rangle_{\rm min}$ is proportional to $I(I+1)$
approximately, and that $\langle \overline{E_{IT}} \rangle_{\rm
min}$ with different $T$ values are different: those with larger
$T$ being systematically higher.  } \figlab{1}
\end{figure}

 As in earlier works, we investigate here the relation
$\langle \overline{E_I} \rangle_{\rm min}  \propto I(I+1)$, where
$\langle \overline{E_I} \rangle_{\rm min}$ is obtained by
averaging $\overline{E_I}$ over the cases of the ensemble in which
$\overline{E_I} ~ (I \sim I_{\rm min})$ is the lowest in energy.
From Fig. 2, one sees that this relation is very robust with
inclusion of the isospin  degree of freedom $T$, i.e., $\langle
\overline{E_{I,T}} \rangle_{\rm min} \propto I(I+1)$. A new and
interesting  observation here is that the  $\langle
\overline{E_{I,T}} \rangle_{\rm min}$ results, which are  obtained
by  averaging over the case in which $\overline{E_{IT}} ~ (I \sim
I_{\rm min}, T \sim T_{\rm  min})$ is the lowest energy, can be
classified according to their  $T$ values: the values of $\langle
\overline{E_{I,T}} \rangle_{\rm min}$ with larger $T$ are
systematically higher.

As we will show later, closer inspection of our calculated results
confirms that $\overline{E_T} = \sum_{I}  (2I+1)
\overline{E_{I,T}} D_{IT} / \left(\sum_I (2I+1)  D_{IT}  \right) =
E_0 + C T(T+1)$ for each individual   run,  which was shown by
French many years ago \cite{French1,French2}.  $D_T=\sum_I
(2I+1)D_{IT}$. Note that the $(2I+1)$ factor is essential  for
proper definition of fixed-$T$ centroids.

Below we discuss fixed-$T$ centroids by propagation equations.
With nucleons occupying say $(j_1, j_2, \ldots)$ orbits, the
spectrum generating algebra is $U({\cal N})$, ${\cal
N}=\sum_i\,2(2j_i+1)$ with the factor 2 appearing due to isospin.
For $n$ nucleons with isospin $T$, the $T$ quantum number labels
the irreducible representations (irreps) of $SU(2)$ algebra that
appears in the direct product (space-isospin) subalgebra $U({\cal
N}/2) \otimes SU(2)$ of $U({\cal N})$. Then the irreps of $U({\cal
N}/2)$ are completely specified by $(n,T)$. A one plus two-body
hamiltonian $H=h(1)+V(2)$, which preserves angular momentum and
isospin is defined by the single particle energies (spe)
$\epsilon_i$ and by the two-body matrix elements
$V^{Jt}_{ijkl}=\langle (kl)Jt \mid V(2) \mid (ij)Jt\rangle$, where
$|(ij)Jt\rangle$ are anti-symmetrized two particle states. With
$H=h(1)+V(2)$, $\overline{E_{T}}$ are polynomials in the scalars
particle number $n$ and $T(T+1)$: $\overline{E_{T}}=a_0+a_1\,n +
a_2 n^2 + a_3 T(T+1)$; see Refs. \cite{French1,French2}.  Solving
for the $a_i$'s using $\overline{E_{(n,T)}}$ for $n \leq 2$, one
obtains the following propagation formula
\begin{eqnarray}
&&  \overline{E_{T}} = n  \langle h(1) \rangle^{1,\frac{1}{2}}
\nonumber \\
&& + \left\{  \frac{n(n+2)}{8} - \frac{T(T+1)}{2} \right\}
\langle V(2)\rangle^{2,0} \nonumber \\
&&+ \left\{ \frac{3n(n-2)}{8} + \frac{T(T+1)}{2}\right\} \langle
V(2)\rangle^{2,1} ~  \label{kota-eq1}
\end{eqnarray}
with
\begin{eqnarray}
 && \langle h(1) \rangle^{1, \frac{1}{2}} =
\frac{1}{\cal N}
 \sum_i  2(2j_i+1) \epsilon_i ~, \nonumber \\
&& \langle V(2)\rangle^{2,t}= \frac{1}{D_t}    \sum_{i \ge j; J}
V^{Jt}_{ijij}(2J+1) ~. \nonumber
\end{eqnarray}
From Eq. (\ref{kota-eq1}), $\overline{E_{T_{\rm
max}}}-\overline{E_{T}} = \frac{1}{2}\{\langle
V(2)\rangle^{2,1}-\langle V(2)\rangle^{2,0}\}\;\{T_{\rm max}(
T_{\rm max}+1) - T(T+1)\}$. With the interaction matrix elements
$V_{ijij}^{J,t}$ chosen to be zero centered independent Gaussian
random variables, ground states will have $T=0$ or $T_{\rm max}$,
with  $50$\% probability for each of them.

\section{energy centroids of spin $I$ states under random
three-body interactions}

The outcome of random  three-body interactions for $sd$ bosons was
first studied by Bijker and Frank in Ref. \cite{Bijker}, where it
was found that the inclusion of random three-body interactions
does not drastically change the pattern of spin $I$ distribution
in the ground states, in comparison with the results calculated by
using random two-body interactions, if boson number $n$ is much
larger than three. In this section we study whether or not the
pattern of $\overline{E_I}$ becomes different if one includes
random three-body interactions. To highlight the feature of
$\overline{E_I}$ with three-body part, we use a Hamiltonian with
{\it only} three-body interactions defined in Appendix B. The
three-body interaction parameters are chosen to be random and
follow the Gaussian distribution.

\begin{figure}[tbp]
\includegraphics[width=8cm,clip]{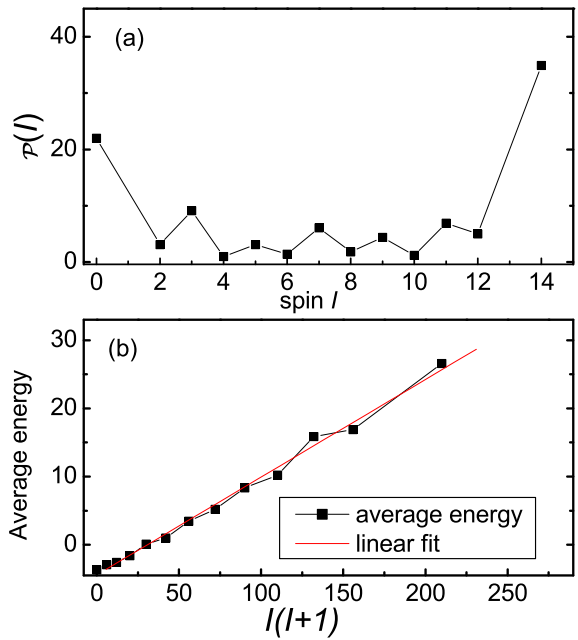}
\caption[fig3]{ (a) ${\cal P}(I)$ with pure random three-body
interactions for $sd$ boson systems.  One sees that ${\cal P}(I)$
is large when $I$ equals $I_{\rm max}$ and $I_{\rm min}$. ${\cal
P}(I)$ also exhibits an odd-even staggering: ${\cal P}(I)$ is
relatively large for odd $I$ values and small for even $I$ values.
(b) $\overline{E_I}$ versus $I(I+1)$. One sees that
$\overline{E_I} \simeq E_0 + C I(I+1)$ for $sd$ bosons with three
body random interactions. In this figure the total boson number
$n$ is 7. Similar results are obtained for $n=6-20$. } \figlab{1}
\end{figure}

Figure 3(a) is a typical example of probability (denoted by ${\cal
P}(I)$) for $\overline{E_I}$ to be the lowest with pure random
three-body interactions of  seven $sd$ boson systems.
Interestingly but as expected, one sees that ${\cal P}(I)$ is
large when $I\sim 0$ or $I\sim I_{\rm max}$. One also sees a very
apparent odd-even staggering of ${\cal P}(I)$ values, i.e., ${\cal
P}(I)$ is large when $I$ is odd and relatively smaller when $I$ is
even.  This behavior was also noticed and discussed in Ref.
\cite{Zhaox} when only the TBRE was used.

Figure 3(b) presents $\langle \overline{E_I} \rangle_{\rm min}$
versus $I(I+1)$ for seven $sd$ bosons.  A linear correlation
between $\langle \overline{E_I} \rangle_{\rm min}$ and $I(I+1)$
can be easily seen.

\begin{figure}[tbp]
\includegraphics[width=8.9cm,clip]{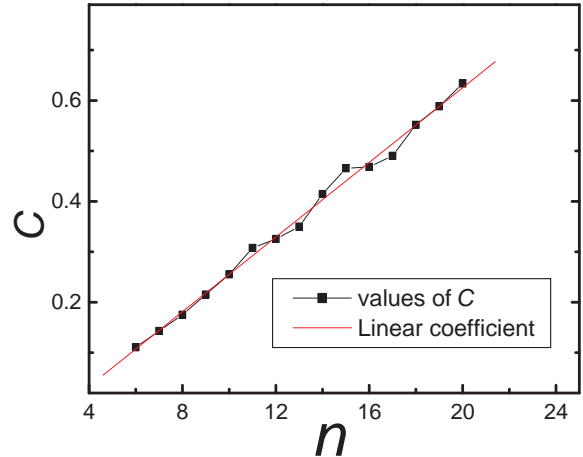}
\caption[fig4]{ The value of $C$ in the   relation $\langle
\overline{E_I} \rangle_{\rm min} \simeq CI(I+1)$ for $sd$-boson
systems versus boson number $n$. We see that the value of $C$ is
approximately proportional to particle number $n$. } \figlab{1}
\end{figure}

A difference between $\langle \overline{E_I} \rangle_{\rm min}$
under the TBRE and that obtained by using random three-body
interactions is found when one studies the correlation $\langle
\overline{E_I} \rangle_{\rm min}=CI(I+1)$ for $sd$-boson systems
with different particle number $n$. For the case by using the
TBRE, the value of coefficient $C$ is not sensitive to particle
number $n$ but sensitive to the single-particle levels in which
the valence particles are active; for systems in which one takes
random three-body  interactions, the situation becomes different.
In Fig. 4 we see that the value of $C$  increases with boson
number $n$ linearly with small fluctuations.

Now let us investigate the energy centroids of spin $I$ states by
using  propagation equations for random three-body interactions.
In order to understand  the particle number dependence seen in
Fig. 4, we consider spin $I$ centroids $\overline{E_{I}}$
generated by random three-body hamiltonians for identical nucleons
in a single-$j$ shell. Firstly $\overline{E_{I}}$'s correspond to
averages over the space defined by the irreps $n$ and $I$ of
$U(2j+1)$ and $SO(3)$ respectively in $U(2j+1) \supset SO(3)$. We
start with the approximate formula (see Eq. (7) in Ref.
\cite{Kota}),
\begin{eqnarray}
&& \overline{E_{I}} = \left[ \langle H(3) \rangle^n -\frac{3}{2}
 \frac{\langle I_z^2 \widetilde{H(3)}\rangle^n}{\langle I_z^2\rangle^n} \right] \nonumber \\
 && + \frac{1}{2}  \frac{\langle I_z^2
\widetilde{H(3)}\rangle^n}{\left[\langle I_z^2\rangle^n\right]^2}
I(I+1) \label{Kotax-4}
\end{eqnarray}
which should be valid for $j>> n>> 3$. In Eq. (\ref{Kotax-4}),
$\langle H(3) \rangle^n$ is the average of $H$ over $n$ particle
space and $\widetilde{H}$ is $H$ with the average part removed
(this is made more clear ahead). Denoting three particle
antisymmetric states by $|(j)^3;\alpha I\rangle$ with $\alpha$
being the extra label required to completely specify the states,
diagonal $3$-particle matrix elements of $H(3)$ are $G_{\alpha J}
= \langle (j)^3;\alpha J \mid H(3) \mid (j)^3;\alpha J\rangle$.

To proceed further it is necessary to consider the tensorial
decomposition of $I^2$ and $H$ operators with respect to $U(2j+1)$
and the tensors are denoted by $\nu$ \cite{French1,French2}. A
general  $k$-body operator will have $\nu=0$, $1$, $\ldots$, $k$
parts. However for a single $j$ shell the $\nu=1$ part will be
zero. Thus $I^2 = (I^2)^{\nu=0} + (I^2)^{\nu=2}$ and
$H(3)=H^{\nu=0}(3) + H^{\nu=2}(3) + H^{\nu=3}(3)$. The $\langle H
\rangle^n$ and $\langle I_z^2\rangle^n=\frac{1}{3} \langle
I^2\rangle^n$ are generated by the $\nu=0$ parts, and
\begin{eqnarray}
&& \langle I_z^2\rangle^n=\frac{1}{3}\langle I^2\rangle^n
\nonumber \\
&&= \frac{1}{6}\, n (2j+1 -n)(j+1) \simeq  \frac{n}{3} j(j+1) ~.
\label{Kotaeq6}
\end{eqnarray}
Note that $\widetilde{H} = H-H^{\nu=0}$, and  $\langle I^2
\widetilde{H(3)}\rangle^n = \langle (I^2)^{\nu=2}
H^{\nu=2}(3)\rangle^n = \langle (I^2)^{\nu=2} H(3)\rangle^n$. It
should be recognized \cite{French1} that $H^{\nu=2}(3) =
(\hat{n}-2) F^{\nu=2}(2)$, where  $F$ is a two-body operator with
rank $\nu=2$ and $\hat{n}$ is number operator. For our purpose, it
is not necessary to know the exact form of the $F$ operator. The
propagation equation for the $n$ particle average $\langle
(I^2)^{\nu=2} F^{\nu=2}(2)\rangle^n$ is well known (see for
example Eq. (A2)  of Ref. \cite{Kota}). Using this we obtain
\begin{eqnarray}
&& \langle (I^2)^{\nu=2} H(3) \rangle^n =
\frac{n(n-1)(n-2)(2j+1-n)(2j-n)}{2(2j-1)(2j-2)} \nonumber \\
&& ~~~~~~~~~ ~~~~~ \times  \langle (I^2)^{\nu=2} F^{\nu=2}(2)
\rangle^2 ~ . \label{Kotax-5}
\end{eqnarray}
Putting $n=3$ on both sides of Eq. (\ref{Kotax-5}), the average
involving $F$ can be eliminated and this gives
\begin{eqnarray}
&& \langle (I^2)^{\nu=2} H(3) \rangle^n =
\frac{n(n-1)(n-2)}{6(2j-2)(2j-3)}  \nonumber \\
&& ~~~~~~~~~ (2j+1-n)(2j-n) \langle (I^2)^{\nu=2} H(3)\rangle^3 ~
.  \label{Kotax-6}
\end{eqnarray}
As given in Ref. \cite{Kota},
$(I^2)^{\nu=2}=I^2-\frac{\hat{n}}{2}(2j+1-\hat{n}) (j+1)$.
Combining Eq. (\ref{Kotax-6}) with Eqs. (\ref{Kotax-4})
 and (\ref{Kotaeq6}) give the final formula for
$\overline{E_{(n,I)}}$ with the $I(I+1)$ term carrying linear $n$
dependence
\begin{eqnarray}
&& \overline{E_{(n,I)}}  \simeq E_0 \nonumber \\
& & + \frac{3n}{2} \left[ \frac{\sum_{\alpha,J} \{J(J+1)-3j(j+1)\}
G_{\alpha J} (2J+1)}{[j(j+1(2j+1)]^2 (2j+1)}\right] \nonumber \\
&& ~~~~~~~~~~~~~~~ \times  I(I+1) ~, \label{Kotaeq9}
\end{eqnarray}
where $E_0$ is a constant determined by $G_{\alpha J}$, $n$ and
$j$. The form of $E_0$ is given in Appendix A. The second term in
Eq. (\ref{Kotaeq9}) clearly gives a linear $n$-dependence for $C$.
This dependence, as seen from the unitary decomposition, comes
from the $\nu=2$ part of $H(3)$ which in turn is responsible for
the $I(I+1)$ term.

Similarly, one investigates  $\overline{E_{(n,I)}}$ versus
$I(I+1)$ for bosons with spin $l$. Eq. (\ref{Kotax-4}) remains the
same but
\begin{eqnarray}
&& \langle I_z^2\rangle^n =  \frac{1}{6} n (2 l+1+n)l \nonumber \\
&& \langle I_z^2 \widetilde{H(3)}\rangle^n  = \nonumber \\
&&\frac{1}{3} \left[
\frac{n(n-1)(n-2)(2l+1+n)(2l+2+n)}{(2l+1)(2l+2)(2l+3)(2l+4)
(2l+5)}\right]  \times \nonumber \\
& & \sum_{\alpha,L} \{L(L+1)-3l(l+2)\} G_{\alpha L} (2L+1) ~.
\end{eqnarray}
Then one obtains
\begin{eqnarray}
&& \overline{E_{(n,I)}}  \simeq E'_0  + n \frac{6}{l^2 (2l+1)(2l+2) \cdots (2l+5)}   \nonumber \\
&& ~~\sum_{\alpha,L} \left[L(L+1)-3l(l+2)\right] G_{\alpha L}
\times  I(I+1) ~, \label{d-boson}
\end{eqnarray}
where $G_{\alpha L}$ are three-body matrix elements for bosons
with spin $l$. For $d$ bosons there are five three-body matrices
with $L=0,2,3,4$ and 6, respectively. By using Eq.(\ref{d-boson}),
one obtains the value of coefficient $C$ in the relation $\langle
\overline{E_I} \rangle_{\rm min} \simeq CI(I+1)$ for $d$ boson
systems:
\begin{eqnarray}
&& C = \sqrt{\frac{2}{\pi}} \frac{6n}{l^2(2l+1)(2l+2) \cdots (2l+5)} \nonumber \\
&& \sqrt{\sum_L \left( (L(L+1)-3l(l+2)) (2L+1) \right)^2}
 \nonumber \\
&&=\frac{\sqrt{997\pi}}{1680} n \simeq 0.03331 n ~,
\end{eqnarray}
where the constant $\sqrt{\frac{2}{\pi}}$ comes from the integral
\begin{eqnarray}
&& \frac{2}{\sqrt{2\pi}}\int^{\infty}_0 x  {\rm
exp}(-\frac{x^2}{2})~ {\rm d} x  ~. \nonumber
\end{eqnarray}
The value of $C$ obtained by our numerical experiments (1000 runs
of the ensemble with random three-body interactions for $d$ boson
systems of $n=6$-20)  is $0.03499n$. The value of $C$ is therefore
reasonably reproduced by the propagation equation (which predicts
$C=0.03331n$ as discussed above).


\section{Energy centroids with fixed irreducible representation}

In this section we consider two interesting  examples: (i) energy
centroids with fixed irreps of the $SU_{sd}(3) \oplus SU_{pf}(3)$
limit of $sdpf$IBM which was mentioned in Ref. \cite{Kota2}; (ii)
energy centroids with fixed Wigner's spin-isospin supermultiplet
$SU(4)$ irreps for $(2s1d)$ shell nuclei.

Let us first consider energy centroids with fixed irreps $[n_{sd}
(\lambda_{sd} \mu_{sd}): n_{pf} (\lambda_{pf} \mu_{pf})]$ of
$[U_{sd}(6) \supset SU_{sd}(3)] \oplus [U_{pf}(10) \supset
SU_{pf}(3)]$ subalgebra of the spectrum generating algebra (SGA)
$U_{sdpf}(16)$ of $sdpf$IBM; see Refs. \cite{Long} for details of
the $SU(3)$ limit of $sdpf$IBM. Given a one plus two-body $sdpf$
hamiltonian, with  boson number $n=n_{sd}+n_{pf}$ where $n_{sd}$
and $n_{pf}$ are number of bosons in $sd$ and $pf$ orbits, the
propagation equation for  energy centroids can be written as
\begin{eqnarray}
 && \overline{E_{n_{sd} (\lambda_{sd} \mu_{sd})\,:\, n_{pf}
(\lambda_{pf} \mu_{pf})}}  \nonumber \\
& = & a_0 + a_1 n_{sd} + a_2 n_{pf} + a_3 \left(
     \begin{array}{c}
     n_{sd} \\
     2 \end{array}   \right)
 + a_4 \left(
     \begin{array}{c}
     n_{pf} \\
     2 \end{array}   \right) \nonumber \\
&& + a_5 n_{sd} n_{pf}  + a_6 C_2(\lambda_{sd} \mu_{sd}) + a_7
C_2(\lambda_{pf} \mu_{pf})\;,  \label{sec4-1}
\end{eqnarray}
where $C_2(\{f\})$ in this section denotes  the eigenvalue of the
quadratic Casimir invariant of a given irrep $\{f\}$. For irreps
$(\lambda \mu)$ of SU(3) it  is given by $C_2(\lambda
\mu)=\lambda^2+\mu^2+\lambda \mu + 3(\lambda+\mu)$. The final
propagation equation  can be obtained by solving for the $a_i$'s
in terms of the centroids with $n \leq 2$. We give this equation
in the Appendix A.

In order to calculate energy centroids using Eq. (\ref{Kota-9}),
the reductions $n_{sd} \rightarrow (\lambda_{sd} \mu_{sd})$ and
$n_{pf} \rightarrow (\lambda_{pf} \mu_{pf})$ are needed. For $sd$
bosons the reductions are well known \cite{Arima0,Iachello} and
for the $pf$ system the reductions are obtained using the method
given in Ref. \cite{Kota-5}.  Following the results for
$sd$IBM-$T$ and $sd$IBM-$ST$ energy centroids given in Ref.
\cite{Kota2}, we consider the basic energy centroids $\langle H
\rangle^{n_{sd} (\lambda_{sd} \mu_{sd}): n_{pf} (\lambda_{pf}
\mu_{pf})}$ with $n \leq 2$ as independent zero centered (with
unit variance) Gaussian random variables, instead of considering
the single-particle energies and two-body matrix elements in
$sdpf$ space as random variables.

With 1000 samples, the probabilities for the energy centroid with
a given $[n_{sd} (\lambda_{sd} \mu_{sd}):  n_{pf} (\lambda_{pf}
\mu_{pf})]$ irrep to be lowest are calculated for boson numbers
$n=8$, $9$ and $10$ and the results are shown in Fig. 5. Firstly
it is seen that  the irreps with the lowest and highest $n_{sd}$
carry most of the probability, about $84$\%. For each of the other
$n_{sd}$ the probability is $1-3$\%. Moreover for the lowest and
highest $n_{sd}$, the probability splits into the lowest and
highest $SU(3)$ irreps. For $n_{sd}=0$ obviously $(\lambda_{sd}
\mu_{sd})=(00)$ and the probability for highest (according to the
eigenvalue of the $SU_{pf}(3)$ quadratic Casimir invariant)
$SU_{pf}(3)$ irreps $(3n,0)$ is $\sim 24$\% and for the lowest
irreps it is $\sim 19$\%. The lowest irreps for $n_{pf}=n=8$, $9$
and $10$ are $(\lambda_{pf} \mu_{pf}) =(00)$, $(30)+(03)$ and
$(00)$, respectively. Similarly, for $n_{sd}=n$, the highest
$SU_{sd}(3)$ irreps are $(2n,0)$ with probability $24$\% and the
lowest irreps are $(02)$, $(00)$ and $(20)$ respectively for
$n_{sd}=n=8$, $9$ and $10$ with probability $\sim 17$\%.  Thus the
results in Fig. 5 show that the energy centroids with the lowest
and highest $[n_{sd} (\lambda_{sd} \mu_{sd}): n_{pf} (\lambda_{pf}
\mu_{pf})]$ irreps carry most of the probability just as with $I$
and $IT$  energy centroids considered in this paper and in many
other examples  considered in Refs. [7-12].

Now let us come to the second example,  energy centroids with
fixed $SU(4)-(ST)$ irreps for the $(2s1d)$ shell model. For
$(2s1d)$ shell nuclei, spin-ispopin supermultiplet $SU(4)$ algebra
appears in the direct product subalgebra $U(6) \otimes SU(4)$ of
$U(24)$ SGA.

We first note that $U(6)$ generates the orbital part and $SU(4)$
generates spin-isospin $(ST)$ quantum numbers via $SU_{ST}(4)
\supset SU_S(2) \otimes SU_T(2)$. For a given number of nucleons
$n$, the allowed $U(4)$ irreps are $\{f\}=\{f_1, f_2, f_3, f_4\}$
with $f_1 \geq f_2 \geq f_3 \geq f_4 \geq 0$, $f_1 \leq 6$ and
$f_1+f_2+f_3+f_4=n$ and the $U(6)$ irreps, by direct product
nature, are $\{\tilde{f}\}$, the transpose of $\{f\}$.  It is
important to note that the equivalent $SU(4)$ irreps are
$\{f_1-f_4, f_2-f_4, f_3-f_4\}$. With these, from now on we will
use $U(4)$ and the irreps $\{f\}$. It is well known that a totally
symmetric $U(4)$ irrep $\{\lambda\} \rightarrow (ST) =
(\frac{\lambda}{2}, \frac{\lambda}{2})$,
$(\frac{\lambda}{2}-1,\frac{\lambda}{2}-1)$, $\ldots$, $(00)$ or
$(\frac{1}{2},\frac{1}{2})$. Using this result and expanding a
given $U(4)$ irrep into totally symmetric $U(4)$ irreps will give
easily $\{f\} \rightarrow (ST)$ reductions. Just as the fixed-$T$
energy centroids propagate, the fixed $\{f\}(ST)$ energy centroids
$\overline{E_{\{f\}(ST)}}$ for a one plus two-body hamiltonian
propagate as the available scalars of maximum body rank 2 are $1$,
$\hat{n}$, $\hat{n}^2$, $C_2(U(4))$, $S^2$ and $T^2$ and the
centroids for $n \leq 2$ are also six in number. The propagation
equation, with $C_2(\{f\})=\sum_i\,f_i^2 +3f_1+f_2-f_3-3f_4$ for
$U(4)$ irrep $\{f\}$, was first discussed in Ref. \cite{Haq} and
we present it in Appendix A, for the sake of convenience.

\begin{figure}
\includegraphics[width=5.0cm]{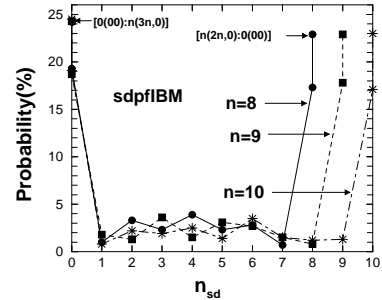}
\caption[fig5]{ Probabilities for $sdpf$IBM centroid  energies
with fixed-$[n_{sd} (\lambda_{sd} \mu_{sd}):n_{pf} (\lambda_{pf}
\mu_{pf})]$ irreps to be lowest in energy vs $n_{sd}$ for systems
of $n=8$, $9$ and $10$ bosons. For $n_{sd}=0$ ($n_{pf}=n$) and
$n_{sd}=n$ ($n_{pf}=0$) the probabilities for $(\lambda_{pf}
\mu_{pf})$ [$(\lambda_{sd} \mu_{sd})=(00)$] and $(\lambda_{sd}
\mu_{sd})$ [$(\lambda_{pf} \mu_{pf})=(00)$] are shown in the
figure. However for $0 < n_{sd} <n$, the  probabilities shown are
the sum of  the probabilities for the $[n_{sd} (\lambda_{sd}
\mu_{sd}):n_{pf} (\lambda_{pf}  \mu_{pf})]$ irreps; for each
$n_{sd}$ and $n_{pf}=n-n_{sd}$ there are  $3-4$ $SU(3)$ irreps but
they are not shown in the figure to avoid clustering of too many
points.  Filled circles, squares and stars are  joined by lines to
guide the eyes. See text for further details. } \figlab{5}
\end{figure}

Just as in the $sdpf$ example, we consider the basic energy
centroids $\langle H \rangle^{\{f\}(ST)}$ with $n \leq 2$ as
independent zero centered (with unit variance) Gaussian random
variables, instead of using $\epsilon_i$ and $V^{Jt}_{ijkl}$ as
random variables, and study the $\{f\}(ST)$ structure of the
ground states.

Using 1000 samples, the probability for a given fixed-$\{f\}(ST)$
energy centroid to be lowest in energy is calculated and the
results are shown in Fig. 6 for $n=8$, $9$, $10$ and $12$. The
probabilities split into three $U(4)$ irreps (other irreps carry
$< 1$\% probability) for  $n=8$, $9$ and $10$, and the
corresponding $(ST)$ values are as shown  in Fig. 6. Energy
centroids with the lowest and highest $U(4)$ irreps carry $\sim
25$\% and $\sim 40$\%, respectively. The lowest irreps are
$\{2^4\}$, $\{32^3\}$ and $\{3^22^2\}$ respectively  for $n=8$,
$9$ and $10$  and  the highest irreps are $\{6,n-6\}$. The third
irreps $\{4^2\}$, $\{54\}$ and $\{5^2\}$,  with probability $\sim
32$\%  for  $n=8$, $9$ and $10$,  are those that carry $S=n/2$ or
$T=n/2$. Besides these, for $n=10$ the irrep $[3^31](00)$ carries
$3.7$\% probability. For the mid-shell example with $n=12$ the
probabilities split into the lowest $\{3^4\}$ and  highest
$\{6^2\}$ irreps with $\sim 25$\% and $\sim 75$\%, respectively.
The lowest irrep supports only $(ST)=(00)$ and the probability for
the highest irrep splits into $\sim 13$\% and $\sim 62$\% for
$(ST)=(00)$ and  $(12,0)+(0,12)$. A very important observation
from Fig. 6 is that the probability for the energy centroid with
lowest $U(4)$ irrep to be lowest is only $\sim 25$\% and it should
be noted that the corresponding $SU(4)$ irreps are $\{0\}(00)$,
$\{1\}(\frac{1}{2} \frac{1}{2})$ and $\{1^2\}(10)+(01)$ for
$n=4k$, $4k+1$ and $4k+2$, respectively, with $k$ being a positive
integer. In fact as discussed in Ref. \cite{Zhao-Scholten},
realistic interactions give ground state wavefunctions having
overlap of $\sim 90$\% with these irreps, i.e. very high
probability for $\alpha$-cluster structure. However detailed
calculations in Ref. \cite{Zhao-Scholten} showed that random
interactions give a very small probability for $\alpha$
clustering. The same  result has been brought out in a simple and
easy manner by the $\{f\}(ST)$ energy centroids.

Finally we point out that the present study extends easily to
$(2p1f)$-shell nuclei by changing the restriction $f_1 \leq 6$ to
$f_1 \leq 10$ in enumerating the $U(4)$ irreps.

\begin{figure}
\includegraphics[width=6.7cm]{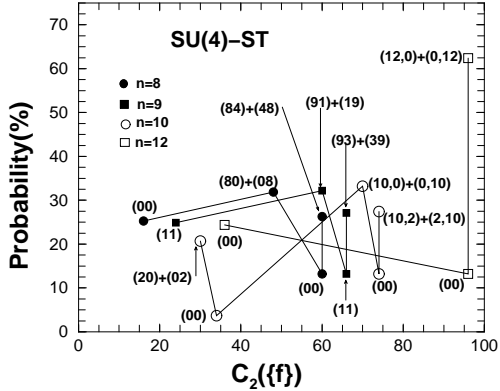}
\caption[fig6]{Probabilities for $(2s1d)$ shell centroid energies
with fixed $\{f\}(ST)$ irreps to be lowest in energy {\it vs}
$C_2(\{f\})$. Results are shown for nucleon numbers $n=8$, $9$,
$10$ and $12$. The irreps carrying probability less than $1$\% are
not shown in the figure. The $U(4)$ irreps $\{f\}$ for the results
in the figure are given in the text. The corresponding $(ST)$
values are shown in the figure as $(2S\,2T)$. Filled circles,
filled squares, open circles and open squares are joined by lines
to guide the eyes. } \figlab{6}
\end{figure}

\section{Discussion and summary}

In this paper we studied the behavior of energy centroids in the
presence of random interactions. First we show  that results of
energy centroids are robust regardless of inclusion of isospin,
based on numerical calculations. We find that $\langle
\overline{E_{IT}} \rangle_{\rm min}$ (and $\langle
\overline{E_{IT}} \rangle_{\rm max}$) values can be classified
according to their $T$ values. The simple relation,
$\overline{E_T} =  E_0 + C T(T+1)$ for each individual  run, is
confirmed and discussed.  We see $\overline{E_{IT}}$ with both
neutrons and protons shows a similar pattern to that of
$\overline{E_{I}}$ with only identical particles discussed in
earlier works.

Second, we find in this paper that for $sd$ boson systems the
feature of $\overline{E_I}$'s is robust with inclusion of random
{\it three-body} interactions. Results of ${\cal P}(I)$, the
probability for $\overline{E_I}$ to be the lowest in energy, and
$\langle \overline{E_I} \rangle_{\rm min}$ (and $\langle
\overline{E_I} \rangle_{\rm max}$), obtained by using the TBRE,
are also applicable to those by using random three-body
interactions, except that the value of coefficient $C$ in the
relation $\langle \overline{E_I} \rangle_{\rm min} \simeq CI(I+1)$
increases linearly with boson number $n$.

Third, such regular patterns can be generalized to energy
centroids with given irreducible representations  of groups  for
boson systems as well as fermion systems. We consider  energy
centroids with fixed irreps of the $SU_{sd}(3) \oplus SU_{pf}(3)$
limit of $sdpf$IBM, and energy centroids with fixed Wigner's
spin-isospin supermultiplet $SU(4)$ irreps for $(2s1d)$ shell
nuclei. We see that the lowest and highest irreps  $\{ f \}$ carry
most of the cases that $\overline{E_I}$ is the lowest in energy,
and  the energy centroids propagate via quadratic Casimir
invariants.

The above results, such as energy centroids with fixed $T$ value,
energy centroids of spin $I$ states under random three-body
interactions, energy centroids of fixed irreps of the IBM models
and shell models, are  discussed by using propagation equations.

Our results suggest that behavior of energy centroids discussed in
Refs. [7-12] is  a {\it very} robust feature for quantum many-body
systems interacting by random forces.

Acknowledgement:  One of the authors (YMZ) would like to thank the
National Natural Science Foundation of China under Grant Nos.
10545001 and 10575070 for supporting this work. Another author
(N. Yoshida) is thankful to the financial support by   ``Academic
Frontier" Project and Organization for Research and  Development
of Innovative Science and Technology (ORDIST) of Kansai
University.

\vspace{0.3in}

\begin{center}

Appendix  {\bf A} ~~  Useful formulas in deriving propagation
equations
\end{center}

First we present the detailed result of $E_0$ in Eq. (9) of Sec.
III.
\begin{eqnarray}
&& E_0 =  \left[ \frac{n^3}{(2j+1)^3}
\sum_{\alpha,J} G_{\alpha J} (2J+1) \right. \nonumber \\
&& ~~~~~~~~
   - \frac{3n^2}{2j(j+1)(2j+1)^3} \nonumber \\
&& ~~~~~~~~ \left. \times \sum_{\alpha,J} \{J(J+1)-3j(j+1)\}
G_{\alpha J} (2J+1) \right] ~.  \nonumber
\end{eqnarray}

Second, we present the propagation equation of energy centroids
with fixed irreps $[n_{sd} (\lambda_{sd} \mu_{sd}): n_{pf}
(\lambda_{pf} \mu_{pf})]$ of $[U_{sd}(6) \supset SU_{sd}(3)]
\oplus [U_{pf}(10) \supset SU_{pf}(3)]$ subalgebra of the spectrum
generating algebra (SGA) $U_{sdpf}(16)$ of $sdpf$ IBM. This is
obtained by solving $a_i$'s in Eq.(\ref{sec4-1}) by centroids with
$n\le 2$.
\begin{eqnarray}
&& \overline{E_{n_{sd} (\lambda_{sd} \mu_{sd})\, :\, n_{pf}
(\lambda_{pf} \mu_{pf})}} \nonumber  \\
&& =\left[1-n_{sd}-n_{pf} \right. \nonumber  \\
&& ~~~ \left. + \left(
     \begin{array}{c}
     n_{sd} \\
     2 \end{array}   \right) +\left(
     \begin{array}{c}
     n_{pf} \\
     2 \end{array}   \right)
+n_{sd} n_{pf} \right] \overline{E_{0(00):0(00)}} \nonumber  \\
&&+ \left[ n_{sd} -2\left(
     \begin{array}{c}
     n_{sd} \\
     2 \end{array}   \right) - n_{sd} n_{pf} \right] \overline{E_{1(20):0(00)}} \nonumber \\
&&  + \left[n_{pf} -2\left(
     \begin{array}{c}
     n_{pf} \\
     2 \end{array}   \right) - n_{sd}
n_{pf} \right]
\overline{E_{0(00):1(30)}} \nonumber \\
&& +  n_{sd} n_{pf}   \overline{E_{1(20):1(30)}}
\nonumber \\
&& + \left[- \frac{5}{9} n_{sd} + \frac{5}{9}\left(
     \begin{array}{c}
     n_{sd} \\
     2 \end{array}   \right) +
 \frac{1}{18} C_2(\lambda_{sd} \mu_{sd}) \right] \overline{E_{2(40):0(00)}} \nonumber \\
&&+ \left[ \frac{5}{9} n_{sd} + \frac{4}{9}\left(
     \begin{array}{c}
     n_{sd} \\
     2 \end{array}   \right) -
\frac{1}{18} C_2(\lambda_{sd} \mu_{sd}) \right] \overline{E_{2(02):0(00)}} \nonumber \\
&& + \left[- \frac{3}{5} n_{pf} + \frac{2}{5}\left(
     \begin{array}{c}
     n_{pf} \\
     2 \end{array}   \right) +
 \frac{1}{30} C_2(\lambda_{pf} \mu_{pf}) \right] \overline{E_{0(00):2(60)}} \nonumber \\
&&+ \left[ \frac{3}{5} n_{pf} + \frac{3}{5}\left(
     \begin{array}{c}
     n_{pf} \\
     2 \end{array}   \right) -
\frac{1}{30} C_2(\lambda_{pf} \mu_{pf}) \right]
\overline{E_{0(00):2(22)}} ~. \nonumber \\  \label{Kota-9}
\end{eqnarray}

Last,  we give the propagation equation for energy centroids with
fixed spin-isospin SU(4) irreps for the $sd$ shell nuclei.  This
was first discussed in Ref. \cite{Haq}.

\begin{eqnarray}
&& \overline{E_{\{f\}(ST)}} \nonumber \\
& =& \left(1-\frac{3}{2}n+\frac{1}{2}
 n^2\right)\,\overline{E_{\{0\}(00)}} +
\left(2n-n^2\right) \overline{E_{\{1\}(\frac{1}{2} \frac{1}{2})}} \nonumber \\
& & + \left[- \frac{9}{8} n + \frac{1}{4} n^2 +  \frac{1}{8}
C_2(\{f\}) +
\frac{1}{4} S(S+1) \right. \nonumber \\
&& ~~~ \left. + \frac{1}{4} T(T+1) \right]\overline{E_{\{2\}(11)}} \nonumber  \\
& & + \left[- \frac{1}{8} n + \frac{1}{8} C_2(\{f\}) -
\frac{1}{4} S(S+1) \right. \nonumber \\
&&  ~~~ \left. - \frac{1}{4} T(T+1) \right] \overline{E_{\{2\}(00)}} \nonumber \\
& & + \left[\frac{3}{8} n +  \frac{1}{8} n^2 - \frac{1}{8}
C_2(\{f\}) + \frac{1}{4} S(S+1) \right. \nonumber \\
&& \left. ~~~ - \frac{1}{4} T(T+1) \right]
\overline{E_{\{1^2\}(10)}} \nonumber \\
& & + \left[ \frac{3}{8} n + \frac{1}{8} n^2 - \frac{1}{8}
C_2(\{f\}) - \frac{1}{4} S(S+1)\right. \nonumber \\
&& \left. ~~~ + \frac{1}{4} T(T+1) \right]
\overline{E_{\{1^2\}(01)}} ~.
\end{eqnarray}

\vspace{0.3in}

\begin{center}
Appendix  {\bf B} ~~ Definition of three-body interactions of $sd$
bosons
\end{center}

In this Appendix we present the definition of the three-body
Hamiltonian of $sd$ boson systems discussed in Sec. III. We note
that the same definition was taken in Ref. \cite{Bijker} by Bijker
and Frank. Discussions of three-body interactions in the
interacting boson model can be found in Ref.\cite{Iachello}.

Our three-body Hamiltonian of $sd$ bosons are given by
\begin{eqnarray}
&& H_3 = \sum_{L=0,2,3,4,6} \sum_{i\le j} \xi_{L_{ij}}
\frac{P^{\dagger}_{L_i} \cdot P_{L_j} + P^{\dagger}_{L_j} \cdot
P_{L_i} }{1+\delta_{ij}}~,
\end{eqnarray}
where
\begin{eqnarray}
 && P_{0_1}^{\dagger} = \frac{1}{6} \left( s^{\dagger} s^{\dagger} s^{\dagger}
 \right)^{(0)}~, P_{0_2}^{\dagger} = \frac{1}{2} \left( s^{\dagger} d^{\dagger} d^{\dagger}
 \right)^{(0)}~, \nonumber \\
 && P_{0_3}^{\dagger} = \frac{1}{6} \left( d^{\dagger} d^{\dagger} d^{\dagger}
 \right)^{(0)}~,  P_{2_1}^{\dagger} = \frac{1}{6} \left( s^{\dagger} s^{\dagger} d^{\dagger}
 \right)^{(2)}~, \nonumber \\
 && P_{2_2}^{\dagger} = \frac{1}{2} \left( s^{\dagger} d^{\dagger} d^{\dagger}
 \right)^{(2)}~, P_{2_3}^{\dagger} = \frac{1}{6} \left( d^{\dagger} d^{\dagger} d^{\dagger}
 \right)^{(2)}~, \nonumber \\
 && P_{3_1}^{\dagger} = \frac{1}{6} \left( d^{\dagger} d^{\dagger} d^{\dagger}
 \right)^{(3)}~, P_{4_1}^{\dagger} = \frac{1}{2} \left( s^{\dagger} d^{\dagger} d^{\dagger}
 \right)^{(4)}~, \nonumber \\
 && P_{4_2}^{\dagger} = \frac{1}{6} \left( d^{\dagger} d^{\dagger} d^{\dagger}
 \right)^{(4)}~, P_{6_1}^{\dagger} = \frac{1}{6} \left( d^{\dagger} d^{\dagger} d^{\dagger}
 \right)^{(6)}~.
\end{eqnarray}
The coefficients $\xi_{L_{ij}}$ are  random and  follow the
Gaussian distribution with their widths given by
\begin{eqnarray}
&& \langle \xi_{L_{ij}} \xi_{L'_{i'j'}} \rangle = (1+\delta_{LL'}
\delta_{ii'} \delta_{jj'})/2 ~.
\end{eqnarray}

\end{document}